\documentclass[prb,byrevtex,floatfix,twocolumn,superscriptaddress]{revtex4-1}
\usepackage{amsmath}
\usepackage{braket}
\usepackage{graphicx}
\usepackage{amssymb,graphicx,afterpage}

\usepackage{txfonts}
\usepackage{multirow}
\usepackage{color}

\begin{document}

\title{Magnetic resonances of multiferroic TbFe$_3$(BO$_3$)$_4$}

\author{D\'avid Szaller}
\affiliation{Department of Physics, Budapest University of Technology and Economics, 1111 Budapest, Hungary}
\affiliation{MTA-BME Lend\"ulet Magneto-optical Spectroscopy Research Group, 1111 Budapest, Hungary}

\author{Vilmos Kocsis}
\affiliation{Department of Physics, Budapest University of Technology and Economics, 1111 Budapest, Hungary}
\affiliation{Condensed Matter Research Group of the
Hungarian Academy of Sciences, 1111 Budapest, Hungary}

\author{S\'andor Bord\'acs}
\affiliation{Department of Physics, Budapest University of Technology and Economics, 1111 Budapest, Hungary}
\affiliation{MTA-BME Lend\"ulet Magneto-optical Spectroscopy Research Group, 1111 Budapest, Hungary}

\author{Titusz Feh\'er}
\affiliation{Department of Physics, Budapest University of Technology and Economics, 1111 Budapest, Hungary}
\affiliation{MTA-BME Lend\"ulet Magneto-optical Spectroscopy Research Group, 1111 Budapest, Hungary}\affiliation{Condensed Matter Research Group of the
Hungarian Academy of Sciences, 1111 Budapest, Hungary}

\author{Toomas R{\~o}{\~o}m}
\affiliation{National Institute of Chemical Physics and Biophysics,
Akadeemia tee 23, 12618 Tallinn, Estonia}

\author{Urmas Nagel}
\affiliation{National Institute of Chemical Physics and Biophysics,
Akadeemia tee 23, 12618 Tallinn, Estonia}

\author{Hans Engelkamp}
\affiliation{High Field Magnet Laboratory (HFML-EMFL), Radboud University, Toernooiveld 7, 6525 ED Nijmegen, The Netherlands}

\author{Kenya Ohgushi}
\affiliation{Institute for Solid State Physics (ISSP), University of Tokyo, Kashiwa-no-ha, Kashiwa, Chiba 277-8561, Japan}

\author{Istv\'an K\'ezsm\'arki}
\affiliation{Department of Physics, Budapest University of Technology and Economics, 1111 Budapest, Hungary}
\affiliation{MTA-BME Lend\"ulet Magneto-optical Spectroscopy Research Group, 1111 Budapest, Hungary}

\date{\today}

\begin{abstract}
Low-energy magnetic excitations of the easy-axis antiferromagnet TbFe$_3$(BO$_3$)$_4$ are investigated by far-infrared absorption and reflection spectroscopy in high magnetic fields up to 30 T. The observed field dependence of the resonance frequencies and the magnetization are reproduced by a mean-field spin model for magnetic fields applied both along and perpendicular to the easy axis. Based on this model we determined the full set of magnetic interactions, including Fe-Fe and Fe-Tb exchange interactions, single-ion anisotropy for Tb ions and $g$-factors, which describe the ground-state spin texture and the low-energy spin excitations of TbFe$_3$(BO$_3$)$_4$. Compared to earlier studies we allow a small canting of the nearly Ising-like Tb moments to achieve a quantitative agreement with the magnetic susceptibility measurements. The additional high energy magnetic resonance lines observed, besides the two resonances expected for a two-sublattice antiferromagnet, suggest a more complex six-sublattice magnetic ground state for TbFe$_3$(BO$_3$)$_4$.
\end{abstract}

\maketitle

\section{Introduction}

Magnetoelectric multiferroics, i.e. materials hosting both ferroelectric and (ferro)magnetic orders, attracted enormous interest due to their potential in  information technology applications.\cite{FreemanSchmid,Fiebig2005,Eerenstein2006,Ramesh2007,Martin2010,Wu2013} 
The magnetoelectric effect emerges not only in the static limit but also in the optical regime, as it is manifested in the difference between the refractive indices of counter-propagating light beams.\cite{Rikken1,Rikken2,Ca2CoSi2O7} Indeed, strong directional dichroism, i.e. different absorption coefficient for light beams travelling in opposite directions, has been reported for spin excitations in multiferroics and was proposed as a new principle of directional light switch operating in the GHz-THz range.\cite{Kezsmarki2011,Bordacs2012,EuYMnO,Takahashi2013,DDPRB2013,Seki,Ca2CoSi2O7,Pimenov}

Recently a new family of magnetoelectric multiferroic crystals, $R$Fe$_3$(BO$_3$)$_4$ rare-earth ferroborates, attracted much attention from the scientific community. Their uniqe crystal structure possessing magnetic iron and rare-earth sites $(R)$ in a chiral arrangement allows the investigation of a wide variety of exotic magnetic and magnetoelectric phenomena.\cite{Zvezdin2005,Zvezdin2006,Kadomtseva2007,Zvezdin2009,Kadomtseva2010,Adem} 

The Fe$^{3+}$ ions are surrounded by edge-sharing O$^{2-}$ octahedra and form quasi one-dimensional helical chains along the trigonal $c$ axis of the crystal.\cite{Joubert,Campa} These helices are expected to be only weakly connected by the Fe-O-O-Fe superexchange paths.\cite{Klimin2005} Thus the magnetic interaction between the iron chains and the rare-earth ions located between the iron helices plays an important role in tuning the effective dimensionality of the magnetic system. The single-ion anisotropy of the rare earth spins is transmitted to the otherwise nearly-isotropic Fe spins via $J_{fd}$ exchange interactions.\cite{Kadomtseva2007} The dominant magnetic interaction is the $J_{dd}$ exchange coupling between Fe spins, which leads to an antiferromagnetic ordering of the Fe subsystem. The rare earth spins---largely separated from each other---remain paramagnetic and are only polarized by the ordered Fe moments.\cite{Pankrats,JPCM2007} 

The strong spin-orbit coupling of the rare-earth ions plays a key role in the multiferroicity of these materials, i.e. their magneto-electric response is dominated by the rare-earth sites.\cite{Zvezdin2005,Zvezdin2006,Kurimaji} Correspondingly, the magnetoelectric properties of rare-earth ferroborates can be efficiently tuned by the selection of different rare earth elements characterized by different magnetic anisotropies.

TbFe$_3$(BO$_3$)$_4$ is a particularly interesting member of the rare-earth ferroborate family. Due to the effect of the strong crystal field, the  ground-state doublet of Tb$^{3+}$ ion is separated from the excited states by a considerable energy gap of $~25\textrm{ meV }(6\textrm{ THz})$.\cite{Popo,CF} Thus at low temperatures the $S_{\textrm{Tb}}=6$ spin of the Tb$^{3+}$ ion behaves like an Ising moment pointing along the trigonal $c$ axis of the crystal. As in the sister compounds $R$=Pr (Ref. [\!\!\citenum{Ritter2010}]) and $R$=Dy (Ref. [\!\!\citenum{Ritter2012}]),
 in TbFe$_3$(BO$_3$)$_4$ the easy-axis anisotropy of the rare-earth ion is transmitted to the antiferromagnetic $S_{\textrm{Fe}}=\frac{5}{2}$ iron system. Consequently, TbFe$_3$(BO$_3$)$_4$ shows a collinear antiferromagnetic order below $T_\textrm{N} = 40\textrm{ K}$ with all spins lying along the $c$ axis.\cite{Hinatsu,JPCM2007,Popo} 

\begin{figure*}[ht!]
\includegraphics[width=7.1in]{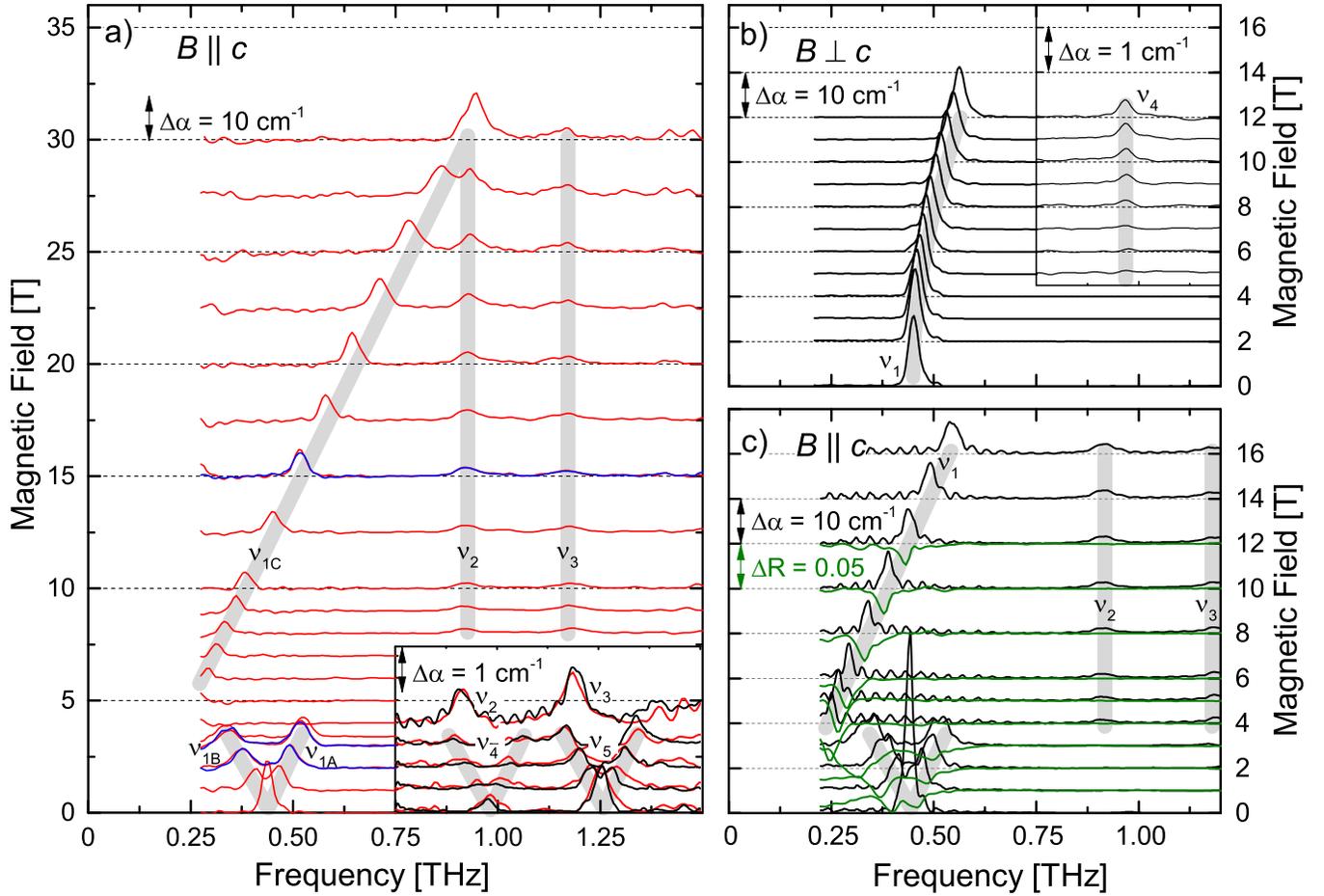}
\caption{(Color online) Magnetic field dependence of the low-frequency magnetic resonances in TbFe$_3$(BO$_3$)$_4$ at $T=2\textrm{ K}$. In panel (a) and (c) the magnetic field points along the trigonal $c$ axis, while in panel (b) the field is perpendicular to the $c$ axis. Light propagation was always parallel or antiparallel to the applied field (Faraday geometry). Magnetic absorption and reflection spectra are vertically shifted in proportion to the magnetic field. In panel (a), red and blue absorption curves measured in HFML correspond to light propagation parallel and antiparallel to the magnetic field, respectively. Black absorption curves on all of the panels and green reflectivity spectra on panel (c) were measured on the TeslaFIR setup. Insets of panels (a) and (b) show the weaker resonances of the corresponding spectra on a 10 times magnified absorption scale. Grey shaded lanes are guides for the eye.}
\label{expfig} 
\end{figure*} 

Magnetic field along the $c$ axis induces spin flop transition and all the Tb moments become parallel to the field while the sublattice magnetization of the antiferromagnetic Fe subsystem rotates to the $ab$ plane and a weak canting develops along the $c$ axis. The spin-flop transition field is $B_{SF}=3.5\textrm{ T}$ at $T=2\textrm{ K}$ and increases with increasing temperature.\cite{JPCM2007} The magnetic order of TbFe$_3$(BO$_3$)$_4$ was widely investigated by both magnetization and elastic neutron scattering experiments.\cite{JPCM2007,Popo} The temperature and field dependent behavior of the static magnetization was reproduced by former mean-field calculations\cite{Popo,CF}. However, the collective magnetic excitations of the ground state were only studied in zero magnetic field via optical spectroscopy.\cite{JETP2011}

Here we investigate the low-energy magnetic excitations of TbFe$_3$(BO$_3$)$_4$ using far-infrared optical spectroscopy up to high magnetic fields applied along and perpendicular to the trigonal $c$ axis. The observed field dependence of the resonance frequencies and the magnetization are reproduced by a mean-field spin model. In contrast to earlier studies in our model we allow a small canting of the quasi-Ising Tb moments and a Tb-Fe exchange interaction is introduced for ions located in the same $ab$ plane to achieve a quantitative agreement with the static and dynamic magnetic properties.

\section{Experimental details}

Fourier transform spectroscopy was used to study the optical absorption and reflection of TbFe$_3$(BO$_3$)$_4$ in the $\nu=0.2 - 2\textrm{ THz}$ frequency range with 8 GHz resolution. The magnetic field dependence of the spectra in the $B=0-17\textrm{ T}$ magnetic field range was investigated using the TeslaFIR setup of the National Institute of Chemical Physics and Biophysics in Tallinn.\cite{Ca2CoSi2O7} 
Optical absorption experiments up to 30 T were carried out in the High Field Magnet Laboratory in Nijmegen (HFML).

The spectra were measured in the Faraday configuration, i.e. in magnetic fields parallel to the direction of light propagation, using oriented single crystal samples with a typical thicknesses of 1 mm. The crystals were grown by a flux method.\citep{JPCM2007,Kurimaji,Hayashida}

\section{Results and discussion}

\subsection{Experimental results}

In Figs. \ref{expfig}(a)-(c), the optical absorption spectrum measured at $T=2\textrm{ K}$ and in $B=0\textrm{ T}$ shows a clear resonance at $\nu_{1}=0.44\textrm{ THz}$. Though this resonance has already been observed and assigned as an antiferromagnetic resonance of the Fe system,\cite{JETP2011} its field dependence has not been investigated so far.


In magnetic fields $B<B_{SF}=3.5\textrm{ T}$ parallel to the $c$ axis the $\nu_{1}$ resonance shows a V-shape splitting to $\nu_{1A}$ and $\nu_{1B}$ modes, as shown in Figs. \ref{expfig}(a) and \ref{expfig}(c). Above $B_{SF}$ these resonances are replaced by a single mode, $\nu_{1C}$, which hardens linearly with increasing field. These resonances are also visible in the field dependence of the reflectivity spectra whose inverse line shape (dip in the reflectivity) as opposed to dielectric resonances supports their magnetic nature. 

Two further resonances with field independent $\nu_{2}=0.93\textrm{ THz}$ and $\nu_{3}=1.17\textrm{ THz}$ frequencies, best visible in Fig. \ref{expfig}(a), also appear in the spin-flop phase. Despite their constant frequency, their oscillator strength grows with increasing field, indicating the magnetic origin of these modes. Besides these pronounced resonances some weaker ones can also be observed in the low-field phase, as shown in the inset of Fig. \ref{expfig}(a). These resonance lines start at $\nu_{4}=0.98\textrm{ THz}$ and $\nu_{5}=1.26\textrm{ THz}$ frequencies in $B=0\textrm{ T}$ and also show a V-shape splitting with increasing magnetic field. The splitting has the same slope as for the $\nu_{1}=0.44\textrm{ THz}$ mode. The weak $\nu_{4}$ and $\nu_{5}$ resonances, and the field independent $\nu_{2}$ and $\nu_{3}$ modes are not visible in the reflectivity spectra of Fig. \ref{expfig}(c).

The absorption coefficient of magnetic excitations is the same for light propagation parallel and antiparallel to the magnetic field within the accuracy of the measurement. It means that despite the chiral crystal structure, the homochiral sample does not show considerable magneto-chiral dichroism (MChD) in the studied frequency window.\cite{Rikken1, Bordacs2012, Ca2CoSi2O7} Static studies revealed that the magnetoelectric effect is mainly associated with Tb sites in TbFe$_3$(BO$_3$)$_4$.\cite{Kurimaji} Since static and dynamic magnetoelectric effects are closely related by a sum rule \cite{KK} and the latters are responsible for the directional dichroism in the THz spectral range, the absence of MChD suggests that the observed excitations are of purely magnetic origin, and belong to the Fe subsystem. This is in accordance with the strong Ising character of the Tb moments, which act on the Fe spins as a static internal magnetic field but otherwise do not contribute to the spin dynamics. In contrast, a recent study\cite{Pimenov} found strong directional dichroism in the easy plane antiferromagnet Sm$_{0.5}$La$_{0.5}$Fe$_3$(BO$_3$)$_4$, which originates from the coupled dynamics of the rare-earth sites with magnetoelectric character and the Fe spins.


As shown in Fig. \ref{expfig}(b), the $\nu_{1}$ resonance exhibits a quadratic shift towards higher frequencies with increasing magnetic field applied perpendicular to the $c$ axis, as expected for an easy axis antiferromagnet. The other excitation discernible in this configuration is the $\nu_{4}$ mode, whose frequency is nearly frequency independent in the range $B=0 - 12\textrm{ T}$ although it gains oscillator strength with increasing magnetic field.

\begin{figure*}[ht!]
\includegraphics[width=7.1in]{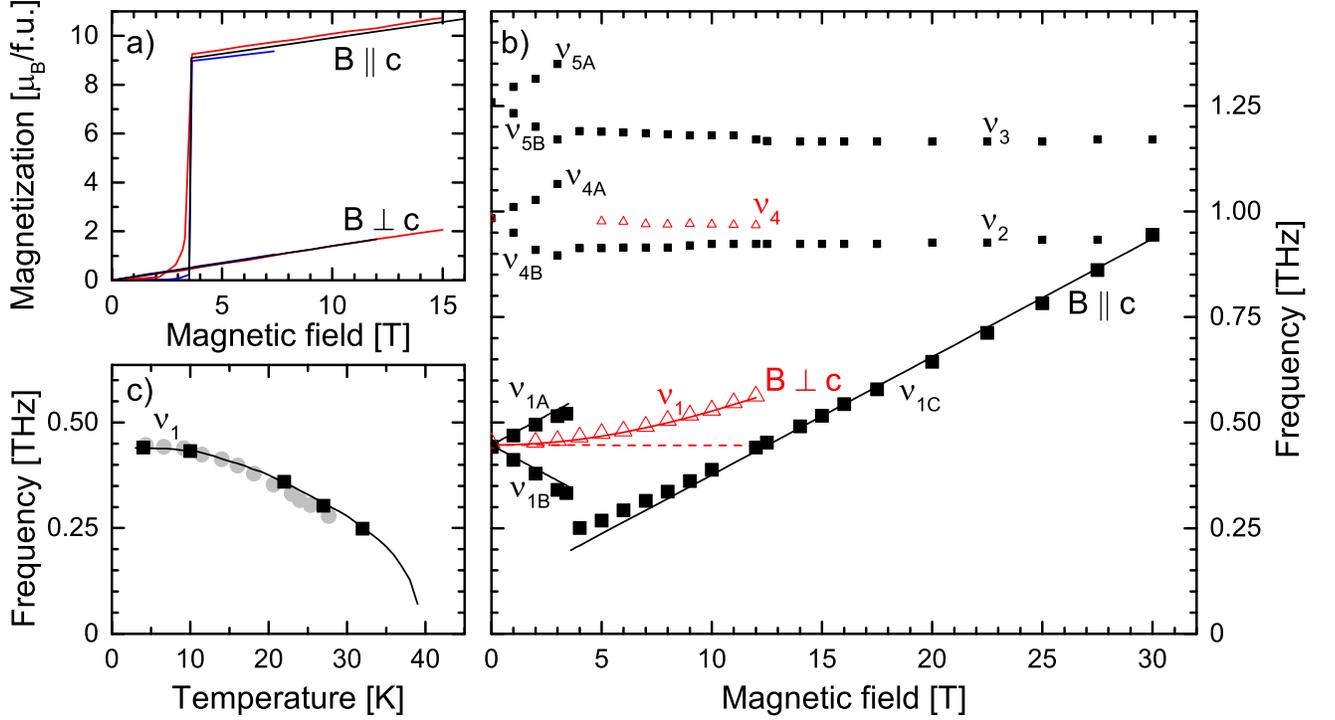}
\caption{(Color online) Comparison of experiments and model calculations. (a) The red and blue curves are experimental magnetization data reproduced from [\!\!\citenum{JPCM2007}] and [\!\!\citenum{Popo}], respectively, while black curves resulted from our classical Monte Carlo simulation. When the magnetic field $\mathbf{B}$ is perpendicular to the trigonal $c$ axis  the three lines coincide within the linewidth. (b) The magnetic field dependence of the low frequency Fe resonances; symbols are used for the experimental data and lines for the simulation results. Black color is used when the magnetic field is parallel and red when it is perpendicular to the trigonal axis. The nearly field-independent dashed mode was not observed in our Faraday geometry measurements (light propagation parallel to the magnetic field), but is allowed in the Voigt geometry (light propagation perpendicular to the magnetic field). Smaller symbols indicate the weaker resonances. (c) The temperature dependence of the zero field resonance, as seen in our optical experiment (black squares), in backward-wave tube experiments\cite{JETP2011} (gray circles), and as calculated (solid black line) from the temperature dependence of the magnetic moment lengths\cite{JPCM2007} using Eq. \ref{omega1}. }
\label{compfig} 
\end{figure*}

The temperature dependence of the $\nu_{1}$ resonance frequency in zero field shows the behavior expected for an anisotropic antiferromagnet, namely it gets softer and broader as the temperature approaches $T_\textrm{N} = 40\textrm{ K}$. Our results measured in transmission and reflection geometries are in good agreement with the previous study of A. M. Kuz’menko et al.,\cite{JETP2011} as shown in Fig. \ref{compfig}(c).

\subsection{Classical mean-field model}

Here we propose a classical mean-field model, which captures the magnetic field dependence of the ground state and that of the strong magnetic modes. We model the Fe and Tb moments, two sublattices for each, as classical vectors with different lengths, $\frac{5}{2}$ for Fe ($\mathbf{S_{FeA}}$ and $\mathbf{S_{FeB}}$) and $6$ for Tb moments ($\mathbf                     {S_{TbA}}$ and $\mathbf{S_{TbB}}$). The energy of a single magnetic unit cell can be written as:

\begin{widetext}
\begin{eqnarray}
H=\Lambda\left(\left(S_{\textrm{TbA}}^c\right)^2+\left(S_{\textrm{TbB}}^c\right)^2\right)-6J_{dd}\mathbf{S_{FeA}S_{FeB}}-6J_{fd1}\left(\mathbf{S_{FeA}S_{TbB}}+\mathbf{S_{FeB}S_{TbA}}\right)-3J_{fd2}\left(\mathbf{S_{FeA}S_{TbA}}+\mathbf{S_{FeB}S_{TbB}}\right) \nonumber \\
-3g_{\textrm{Fe}}\mu_{B}\mathbf{B}\left(\mathbf{S_{FeA}}+\mathbf{S_{FeB}}\right)-g_{\textrm{Tb}}\mu_{B}\mathbf{B}\left(\mathbf{S_{TbA}}+\mathbf{S_{TbB}}\right).
\label{model}
\end{eqnarray}
\end{widetext}

We include a strong negative uniaxial single-ion anisotropy term $\Lambda$ for the Tb sites to model their Ising-like nature. In case of the terms describing the exchange energy contributions, numeric prefactors correspond to the coordination numbers. An isotropic negative $J_{dd}$ exchange term connecting the Fe sublattices is responsible for the antiferromagnetic order. It is reasonable to consider the magnetic anisotropy of the rare-earth system only since in non-magnetic (YFe$_3$(BO$_3$)$_4$) and magnetically isotropic (GdFe$_3$(BO$_3$)$_4$) members of the crystal family the Fe spin system is nearly isotropic Heisenberg-like.\cite{Pankrats2}

The Tb and Fe moments are coupled by the $J_{fd1}$ and $J_{fd2}$ exchange terms, where the dominant $J_{fd1}$ connects  moments on adjacent $ab$ layers and the weaker $J_{fd2}$ links ions on the same $ab$ layers. This kind of coupling was neglected so far,\cite{Popo,Zvezdin2009,CF} since the possible superexchange path corresponding to $J_{fd2}$ is quite long. 

The last two terms of Eq. \ref{model} describe the Zeeman energy, where $\mathbf{B}$ is the external magnetic field.


\subsubsection{Estimation of model parameters based on magnetization data}

Approximate values for the parameters in Eq. \ref{model} can be deducted from the static magnetization data and can be further tuned to fit the lowest excitation frequencies of the system. For the first estimation of the exchange and anisotropy constants we use the Land\'e values $g_{\textrm{Fe}}=2$ and $g_{\textrm{Tb}}=1.5$ given for the free ions, which could be modified by crystal-field effects. 

The slope of the magnetization in $B>B_{SF}$ magnetic field parallel to the $c$ axis is governed by the susceptibility of the Fe subsystem, because in the spin flop phase all the Tb moments point along the magnetic field and do not contribute to the susceptibility. The susceptibility per formula unit $\chi^{c}=0.12\pm 0.01\frac{\mu_{B}}{\textrm{T}}$ is determined by the $J_{dd}$ exchange, hence $J_{dd}\approx-\frac{3g_{\textrm{Fe}}}{2\chi^{c}}=-2.9\pm 0.3\textrm{ meV}$.\cite{Popo,JPCM2007} This value of $J_{dd}$ is in good agreement with most of the previous studies ($J_{dd}\approx-2.1\textrm{ meV}$, $-3.3\textrm{ meV}$, $-3\textrm{ meV}$).\cite{Popo,JPCM2007,CF}


\begin{table*}        
\caption{Parameters of Eq. \ref{model} obtained by using the static magnetization data\cite{Popo,JPCM2007} (first row) and refined parameters using both magnetization and magnetic resonance data (second row). In the last column upper and lower error bounds are indicated by $+$ and $-$, respectively.}
\begin{tabular}{| l | l | l | l | l | l |}
\hline
	$g_{\textrm{Fe}}$	&	$g_{\textrm{Tb}}$	&	$J_{dd} [\textrm{meV}]$	&	$J_{fd1} [\textrm{meV}]$&	$J_{fd2} [\textrm{meV}]$	&	$\Lambda [\textrm{meV}]$		\\
\hline
			2	&	$1.405\pm 0.025$	&	$-2.9 \pm 0.26$			&	$0.04\pm 0.0013$	&	0	&	$-4.1\substack{+1.6 \\ -6.2}$		\\
\hline			
			2	&	$1.365\pm 0.025$		&	$-2.67 \pm 0.15$	&	$0.054\pm 0.004$	&	$0.026\pm 0.009$	&	$-8.1\substack{+3.6 \\ -30}$	\\
\hline
\end{tabular} 
\label{partabl}
\end{table*}

At the spin-flop transition the $\mathbf{S_{TbB}}$ moments occuping one half of the Tb sites flip from $S_{\textrm{TbB}}^c=-6$ to $S_{\textrm{TbB}}^c=6$, giving rise to a $6g_{\textrm{Tb}}$  jump in the magnetization per formula unit. In addition, the Fe moments show some canting in the spin-flop phase which also gives a minor contribution to the jump of the magnetization. The spin-flop transition takes place when the energy of the collinear and spin flop phases are equal, resulting in an approximate expression for the Tb-Fe coupling:  
%
\begin{equation}
\label{j}
J_{fd1}-J_{fd2}/2\approx\frac{g_{\textrm{Tb}} \mu_{B}B_{SF}}{15} + \frac{\left(5+4g_{\textrm{Tb}}\right)\mu_{B}^2 B_{SF}^2}{900}\chi^c,  
\end{equation}
where we neglected higher order terms in $\chi^c$. Due to the different exchange paths $\left\lvert J_{fd2}/2\right\rvert$ is expected to be much smaller than $\left\lvert J_{fd1}\right\rvert$. Correspondingly all of the previous studies neglected the contribution of $J_{fd2}$ and attributed the whole Tb-Fe coupling to the $J_{fd1}$ exchange.\cite{Popo,CF} On the other hand, static magnetization data are not sufficient to unambiguously determine both $J_{fd1}$ and $J_{fd2}$, thus, in this section we assume $J_{fd2}= 0$ following previous works.

The effective field along the $c$ axis acting on the antiferromagnetic Fe system can be approximated by the sum of the external field and the effective field of the Tb moments. Since in the spin-flop phase both of the Tb sublattices are in the $S_{\textrm{Tb}}^c=+6$ state, their effective field on the Fe site is $\frac{6}{g_{\textrm{Fe}}\mu_{B}}\left(J_{fd1}+J_{fd2}/2\right)\approx \frac{6}{g_{\textrm{Fe}}\mu_{B}}J_{fd1}= \frac{4g_{\textrm{Tb}}}{5g_{\textrm{Fe}}}B_{SF} =2.1\textrm{ T}$. Thus, the $\Delta M^c=9.1\pm 0.1\mu_{B}$ magnetization jump\cite{Popo,JPCM2007} at the spin-flop transition is
\begin{subequations}
\begin{eqnarray}
\Delta M^c\left(B_{SF} \right) & = & 6g_{\textrm{Tb}} + \chi^c\left(B_{SF} + \frac{6\left( J_{fd1}+J_{fd2}/2 \right)}{g_{\textrm{Fe}}\mu_{B}}\right) \label{M_egzact} \\
 &\approx& 6g_{\textrm{Tb}} + \chi^c B_{SF}\left(1 +  \frac{4g_{\textrm{Tb}}}{5g_{\textrm{Fe}}}  \right), \label{M_approx}
\end{eqnarray}  
\end{subequations}
giving rise to a refined value of $g_{\textrm{Tb}}\approx 1.405\pm 0.025$ which is significantly lower than the $g_{\textrm{Tb}}=1.5$ Land\'e value. Neutron scattering studies\cite{JPCM2007} reported $g_{\textrm{Tb}}S_{\textrm{Tb}}^c=8.53\mu_{B}$ ordered Tb moment at $T=2\textrm{ K}$, corresponding to $g_{\textrm{Tb}}=1.42$, which is in good accordance with our analysis. 

Using Eq. \ref{j} the strength of the Tb-Fe exchange can be determined, $J_{fd1} \approx 0.04\pm 0.0013\textrm{ meV}$. This is in good agreement with the  $0.044\textrm{ meV}$ and $0.039\textrm{ meV}$ values of previous magnetization studies.\cite{Popo,Zvezdin2009} The same coupling constant was determined from the splitting of the ground quasi-doublet of the Tb ions, which was observed as a splitting of the infrared transitions, corresponding to $0.045\textrm{ meV}$.\cite{CF} 

For fields perpendicular to the $c$ axis, the susceptibility of the system is $\chi^{ab}=0.14\pm 0.002\frac{\mu_{B}}{\textrm{T}}$, which is about $20 \%$ larger than $\chi^{c}$.\cite{Popo,JPCM2007} As the Fe system is expected to be isotropic, the anisotropy of the susceptibility indicates the small canting of the Tb moments and thus can be used to estimate the anisotropy of the Tb sites: $\Lambda\approx-\frac{g_{\textrm{Tb}}}{2(\chi^{ab}-\chi^{c})}=-4.1\substack{+1.6 \\ -6.2}\textrm{ meV}$ (upper and lower error bounds are indicated by $+$ and $-$, respectively). The uncertainty of $\Lambda$ comes from the variation of the experimental values for $\chi^{ab}$ and $\chi^{c}$. However, due to the length of the Tb moments, $\Lambda\left(S_{\textrm{Tb}}^c\right)^2$ gives the dominant energy scale of the system in the studied magnetic field range. This justifies the approximation that Tb moments behave almost like Ising spins. The values obtained for $\Lambda$ correspond to the range of the lowest excited crystal field energy levels calculated for the $\textrm{Tb}^{3+}$ ion.\cite{CF}

The model parameter set obtained above is presented in the first row of Table \ref{partabl}, which reproduces the static magnetization data. However, the static magnetization data only supports a rough estimation of the model parameters. Moreover, in the former expressions only one combination of the two types of Tb-Fe coupling appears, namely $J_{fd1}-J_{fd2}/2$, and therefore in studies based on the magnetization data the minor $J_{fd2}$ was simply neglected.\cite{Popo,CF} In contrast, the magnetic field dependence of the dominant low-frequency magnetic excitations allows us to separate $J_{fd1}$ and $J_{fd2}$ and refine the values of all parameters in the Hamiltonian in Eq. \ref{model}.  

\subsubsection{Determination of model parameters based on magnetic resonances}


Assuming Ising-like Tb moments $\left(\Lambda \rightarrow -\infty\right)$, the zero temperature resonance frequencies of the Fe system can be calculated\cite{Turov} using the $S_{\textrm{Tb}}=6$ and $S_{\textrm{Fe}}=\frac{5}{2}$ values:
\begin{widetext}        
\begin{eqnarray}
\nu_{1}\left( B=0\right) & = & \sqrt{ \left(\left(J_{fd1}-\frac{J_{fd2}}{2}\right) S_{\textrm{Tb}} \right)^2 - 2 J_{dd} S_{\textrm{Fe}}\left(J_{fd1}-\frac{J_{fd2}}{2}\right) S_{\textrm{Tb}} } \label{omega1}\\
\nu_{1A/B}\left( B=B^c <B_{SF}\right) & = & \sqrt{ \left(\left(J_{fd1}-\frac{J_{fd2}}{2}\right) S_{\textrm{Tb}} \right)^2 - 2 J_{dd} S_{\textrm{Fe}}\left(J_{fd1}-\frac{J_{fd2}}{2}\right) S_{\textrm{Tb}} }  \pm g_{\textrm{Fe}}\mu_{B}B \label{omega1AB}\\
\nu_{1C}\left( B=B^c >B_{SF}\right) & = & \left(J_{fd1}+\frac{J_{fd2}}{2}\right) S_{\textrm{Tb}} + g_{\textrm{Fe}}\mu_{B}B \label{omega1C}\\
\nu_{1}\left(  B=B^{ab} \right) & = &  \sqrt{ \left(\left(J_{fd1}-\frac{J_{fd2}}{2}\right) S_{\textrm{Tb}} \right)^2 - 2 J_{dd} S_{\textrm{Fe}}\left(J_{fd1}-\frac{J_{fd2}}{2}\right) S_{\textrm{Tb}} + \left(g_{\textrm{Fe}}\mu_{B}B\right)^2 }.\label{omega1plane}
\end{eqnarray}
\end{widetext}
Using Eq. \ref{omega1}, the $J_{dd}$ Fe-Fe exchange can be determined with higher accuracy than our previous estimation from the $\chi^{c}$ magnetic susceptibility. Based on the experimental $\nu_{1}\left( B=0\right)=0.442\pm 0.005\textrm{ THz}$ frequency value we get $J_{dd}=-2.67\pm 0.15\textrm{ meV}$. 

In the spin flop phase at $B^{c}=B_{SF}$ the effective magnetic field acting on the antiferromagnetic Fe system is $B_{SF}+\frac{6}{g_{\textrm{Fe}}\mu_{B}}\left(J_{fd1}+J_{fd2}/2\right)$, giving information about the sum of $J_{fd1}$ and $J_{fd2}$, thus can be used to unambiguously determine $J_{fd2}$. The experimental value of the $\nu_{1C}$ resonance frequency in the flop phase can be extrapolated to $\nu_{1C}\left(B^{c}=B_{SF}\right)=195\textrm{ GHz}$, corresponding to an effective field of $7\textrm{ T}$. This results in $J_{fd1}=0.054\textrm{ meV}$ and $J_{fd2}=0.026\textrm{ meV}$, and refines the Tb $g$-factor to $g_{\textrm{Tb}}=1.365$ according to Eq. \ref{M_egzact}. The $J_{fd2}$ exchange is indeed weaker than $J_{fd1}$ but does not have a ferromagnetic character, in contrast to former expectations based on the crystal and magnetic structure.\cite{JPCM2007} Thus, in the zero-field ground state the bond corresponding to $J_{fd2}$ is frustrated. 

According to Eqs. \ref{omega1AB} and \ref{omega1C}, the slope of the $\nu_{1A/B}$ and $\nu_{1C}$ modes yields the $g$-factor of the Fe system, which within the error of the measurement is equal to the spin-only $g=2$ value. Using Eq. \ref{omega1plane} to fit the resonance frequencies measured in the $B\perp c$ case we get the same $g$-factor, thus the spin-only Fe $g$-factor is isotropic, as expected. 

When considering the finite temperature excitations of the system, in the zero field case Eq. \ref{omega1} remains valid, only the temperature dependence of the lengths of the $S_{\textrm{Tb}}$ and $S_{\textrm{Fe}}$ ordered moments needs to be to be taken into account. However, a mean-field model like Eq. \ref{model} is not able to properly describe the temperature dependence of the magnetic properties due to neglected thermal fluctuations, thus for a quantitative description additional experimental input is needed. In the elastic neutron scattering studies\cite{JPCM2007} the lengths of the $S_{\textrm{Tb}}$ and $S_{\textrm{Fe}}$ ordered moments were reported in the whole temperature range of the antiferromagnetic phase. By substituting these temperature dependent ordered moments into Eq. \ref{omega1} the temperature dependence of the zero field resonance can be well reproduced with $g_{\textrm{Fe}}=2$, $J_{dd}=-2.67\textrm{ meV}$, $J_{fd1}=0.054\textrm{ meV}$ and $J_{fd2}=0.026\textrm{ meV}$, as shown in Fig. \ref{compfig}(c). Earlier backward-wave oscillator spectroscopy studies\cite{JETP2011} reported the same temperature dependence. 

For finite values of the Tb single-ion anisotropy $\Lambda$, the analytical solution corresponding to Eqs. \ref{omega1}-\ref{omega1plane} is too complicated. Thus we calculated the field dependence of the zero temperature resonances numerically. We used a classical Monte Carlo approach to find the minimal energy configuration of the four-spin system, and determined the resonances by calculating the response to small perturbations. 
The Tb single-ion anisotropy was set to  
\begin{equation}
\Lambda\approx-\frac{g_{\textrm{Tb}}}{2(\chi^{ab}+\frac{3g_{\textrm{Fe}}}{2J_{dd}})}=-8.1\textrm{ meV},
\end{equation}
and the other exchange parameters used in the simulation are listed in the second row of Table \ref{partabl}. The calculated field dependence of the magnetization and antiferromagnetic resonance frequencies reproduce the experimental curves, as shown in Fig. \ref{compfig}(a) and \ref{compfig}(b). With finite $\Lambda$ the Tb moments are not static any more but oscillate with a zero-field resonance frequency of $\nu_{\textrm{Tb}}=\Lambda S_{\textrm{Tb}}=50\textrm{ meV}$, which agrees well with the frequency range of the lowest excited crystal field energy levels calculated for the $\textrm{Tb}^{3+}$ ion.\cite{CF} 

The field independent $\nu_{2}$ and $\nu_{3}$ resonances and the weak $\nu_{4}$ and $\nu_{5}$ modes cannot be explained by this simple classical four-sublattice mean-field spin model. Their presence shows that the Fe sites are crystallographically not equivalent, as is expected for the low temperature $P3_{1}21$ space group\cite{Klimin2005,Fausti} of TbFe$_3$(BO$_3$)$_4$. Thus the proper description of the magnetic resonances is possible only  with six magnetic Fe sublattices which are connected by the various, non-equivalent Fe-O-Fe intrachain and Fe-O-B-O-Fe interchain superexchange paths. Distinction between intrachain and interchain coupling would allow the tuning of the dimensionality of the system, thus the Monte Carlo approach could probably also reproduce the magnetic properties and resonance frequencies at finite temperatures. Nevertheless, due to the weak structural distortion from the room-temperature $R32$ structure to the low temperature $P3_{1}21$, the magnetic properties can be approximated by assuming crystallographically equivalent Fe sites. 

\section{Summary}

In this study we have investigated the low frequency magnetic excitations of the multiferroic TbFe$_3$(BO$_3$)$_4$ using far-infrared spectroscopy. 
We developed a classical mean-field spin model which quantitatively describes the main features in field dependence of the magnetization data\cite{Popo,JPCM2007} and that of the resonance frequencies with a minimal set of magnetic interactions including exchange couplings and single-ion anisotropy. Our far-infrared experiments also pointed out that the magnetic structure of TbFe$_3$(BO$_3$)$_4$ is more complicated than previously expected. There are six inequivalent magnetic Fe sublattices, thus a more detailed neutron diffraction study is necessary to clarify the real magnetic ground state.   

\section{Acknowledgment}

The authors are grateful for enlightening discussions with K. Penc and L. Mih\'aly. This project was funded by Hungarian Research Funds OTKA K108918, PD 111756, K107228 and Bolyai 00565/14/11, by the institutional research funding IUT23-3 of the Estonian Ministry of Education and Research and the European Regional Development Fund project TK134 (TR and UN). We acknowledge the support of the HFML-RU/FOM, member of the European Magnetic Field Laboratory. D. Sz. was supported by the  \'UNKP-16-3/III. New National Excellence Program of the Ministry of Human Capacities.


\begin{references}

\bibitem{FreemanSchmid}A. J. Freeman and H. Schmid (eds.) \textit{Magnetoelectric interaction phenomena in crystals} (Gordon and Breach, London, 1995).
\bibitem{Fiebig2005}M. Fiebig,
J. Phys. D: Appl. Phys. {\bf 38}, R123 (2005).
\bibitem{Eerenstein2006}W. Eerenstein, N. D. Mathur and J. F. Scott,
Nature {\bf 44}, 759 (2006).
\bibitem{Ramesh2007}R. Ramesh and N. A. Spaldin,
Nat. Mater. {\bf 6}, 7 (2007).
\bibitem{Martin2010}L. W. Martin, Y.-H. Chuc and R. Ramesh,
Materials Science and Engineering R {\bf 68}, 89 (2010).
\bibitem{Wu2013} S. M. Wu, Shane A. Cybart, D. Yi, James M. Parker, R. Ramesh and R. C.
Dynes,
Phys. Rev. Lett. {\bf 110}, 067202 (2013).
\bibitem{Ca2CoSi2O7}I. K\'ezsm\'arki, D. Szaller, S. Bord\'acs, V. Kocsis, Y. Tokunaga, Y. Taguchi, H. Murakawa, Y. Tokura, H. Engelkamp, T. R\~o\~om and U. Nagel,
Nat. Commun. {\bf 5}, 3203 (2013).
\bibitem{Rikken1} G. L. J. A. Rikken and E. Raupach,
Nature {\bf 390}, 493 (1997).
\bibitem{Rikken2} G. L. J. A. Rikken, C. Strohm and P. Wyder,
Phys. Rev. Lett. {\bf 89}, 133005 (2002).
\bibitem{Kezsmarki2011}I. K\'ezsm\'arki, N. Kida, H. Murakawa, S. Bord\'acs, Y. Onose, and Y. Tokura,
Phys. Rev. Lett. {\bf 106}, 057403 (2011).
\bibitem{Bordacs2012}S. Bord\'acs, I. K\'ezsm\'arki, D. Szaller, L. Demk\'o, N. Kida, H. Murakawa, Y. Onose, R. Shimano, T. R\~o\~om, U. Nagel, S. Miyahara, N. Furukawa and Y. Tokura,
Nat. Phys. {\bf 8}, 734 (2012).
\bibitem{DDPRB2013}D. Szaller, S. Bord\'acs and I. K\'ezsm\'arki,
Phys. Rev. B {\bf 87}, 014421 (2013).
\bibitem{EuYMnO}Y. Takahashi, R. Shimano, Y. Kaneko, H. Murakawa and Y.
Tokura,
Nat. Phys. {\bf 8}, 121 (2012).
\bibitem{Takahashi2013}Y. Takahashi, Y. Yamasaki, and Y. Tokura,
Phys. Rev. Lett. {\bf 111}, 037204 (2013).
\bibitem{Seki}Y. Okamura, F. Kagawa,	M. Mochizuki,	M. Kubota,	S. Seki,	S. Ishiwata,	M. Kawasaki,	Y. Onose	and Y. Tokura,
Nat. Commun. {\bf 4}, 2391 (2013).	
\bibitem{Pimenov}A. M. Kuzmenko, V. Dziom, A. Shuvaev, Anna Pimenov, M. Schiebl, A. A. Mukhin, V. Yu. Ivanov, I. A. Gudim, L. N. Bezmaternykh and A. Pimenov,
Phys. Rev. B {\bf 92}, 184409 (2015).
\bibitem{Zvezdin2009}A. K. Zvezdin, A. M. Kadomtseva, Yu. F. Popov, G. P. Vorob’ev, A. P. Pyatakov, V. Yu. Ivanov, A. M. Kuz’menko, A. A. Mukhin, L. N. Bezmaternykh and I. A. Gudim,
JETP {\bf 109}, 68 (2009).
\bibitem{Zvezdin2005}A. K. Zvezdin, S. S. Krotov, A. M. Kadomtseva, G. P. Vorob’ev, Yu. F. Popov, A. P. Pyatakov, L. N. Bezmaternykh and E. A. Popova, 
JETP Lett. {\bf 81}, 272 (2005).
\bibitem{Zvezdin2006}A. K. Zvezdin, G. P. Vorob’ev, A. M. Kadomtseva, Yu. F. Popov, A. P. Pyatakov, L. N. Bezmaternykh, A. V. Kuvardin and E. A. Popova, 
JETP Lett. {\bf 83}, 509 (2006).
\bibitem{Kadomtseva2007}A. M. Kadomtseva, A. K. Zvezdin, A. P. Pyatakov, A. V. Kuvardin, G. P. Vorob’ev, Yu. F. Popov and L. N. Bezmaternykh, 
JETP {\bf 105}, 116 (2007).
\bibitem{Kadomtseva2010}A. M. Kadomtseva, Yu. F. Popov, G. P. Vorob’ev, A. P. Pyatakov, S. S. Krotov and K. I. Kamilov, Low Temp. Phys. {\bf 36}, 511 (2010).
\bibitem{Adem}U. Adem, L. Wang, D. Fausti, W. Schottenhamel, P. H. M. van Loosdrecht, A. Vasiliev, L. N. Bezmaternykh, B. Buchner, C. Hess and R. Klingeler, 
Phys. Rev. B  {\bf 82}, 064406 (2010).
\bibitem{Joubert}J. C. Joubert, W. B. White and R. Roy, 
J. Appl. Cryst. {\bf 1}, 318 (1968).
\bibitem{Campa}J. A. Camp\'a, C. Cascales, E. Guti\'errez-Puebla, M. A. Monge, I. Rasines and C. Ru\'ız-Valero,
Chem. Mater. {\bf 9}, 237 (1997).
\bibitem{Klimin2005}S. A. Klimin, D. Fausti, A. Meetsma, L. N. Bezmaternykh, P. H. M. van Loosdrecht and T. T. M. Palstra,
Acta Cryst. B \textbf{61}, 481 (2005).
\bibitem{JPCM2007}C. Ritter, A. Balaev, A. Vorotynov, G. Petrakovskii, D. Velikanov, V. Temerov, and I. Gudim, 
J. Phys.: Condens. Matter {\bf 19}, 196227 (2007).
\bibitem{Pankrats}A. I. Pankrats, G. A. Petrakovskii, L. N. Bezmaternykh and O. A. Bayukov,
JETP {\bf 99} 766 (2004).
\bibitem{Kurimaji}T. Kurumaji, K. Ohgushi, and Y. Tokura,
Phys. Rev. B {\bf 89}, 195126 (2014).
\bibitem{Popo}E. A. Popova, D. V. Volkov, A. N. Vasiliev, A. A. Demidov, N. P. Kolmakova, I. A. Gudim, L. N. Bezmaternykh, N. Tristan, Yu. Skourski, B. B\"uchner, C. Hess, and R. Klingeler,
Phys. Rev. B {\bf 75}, 224413 (2007).
\bibitem{CF}M. N. Popova, T. N. Stanislavchuk, B. Z. Malkin and L. N. Bezmaternykh,
J. Phys.: Condens. Matter {\bf 24} 196002 (2012).
\bibitem{Ritter2010} C. Ritter, A. Vorotynov, A. I. Pankrats, G. A. Petrakovskii, V. Temerov, I. Gudim and R. Szymczak, 
J. Phys.: Condens. Matter {\bf 22} 206002 (2010).
\bibitem{Ritter2012}C. Ritter, A. I. Pankrats, I. Gudim and A Vorotynov,
J. Phys.: Conf. Ser. {\bf 340} 012065 (2012).
\bibitem{Hinatsu}Y. Hinatsu, Y. Doi, K. Ito, M. Wakeshima and A. Alemi,
J. Solid State Chem. {\bf 172} 438 (2003).
\bibitem{JETP2011}A. M. Kuz’menko, A. A. Mukhin, V. Yu. Ivanov, A. M. Kadomtseva, S. P. Lebedeva, and L. N. Bezmaternykh,
JETP {\bf 113}, 113 (2011).
\bibitem{Hayashida}S. Hayashida, M. Soda, S. Itoh, T. Yokoo, K. Ohgushi, D. Kawana, H. M. R\o nnow and T. Masuda
Phys. Rev. B {\bf 92}, 054402 (2015).
\bibitem{KK}D. Szaller, S. Bord\'acs, V. Kocsis, T. R\~o\~om, U. Nagel and I. K\'ezsm\'arki,
Phys. Rev. B {\bf 89}, 184419 (2014).
\bibitem{Pankrats2}A. I. Pankrats, G. A. PetrakovskiI, L. N. Bezmaternykh and V. L. Temerov,
Phys. Solid State {\bf 50} 79 (2008).
\bibitem{Turov}E. A. Turov, \textit{Physical properties of magnetically ordered crystals}, Moscow: Izdat. Acad. Sci. SSSR, (1963). 
\bibitem{Fausti}D. Fausti, A. A. Nugroho, P. H. M. van Loosdrecht, S. A. Klimin, M. N. Popova, and L. N. Bezmaternykh,
Phys. Rev. B {\bf 74}, 024403 (2006).


\end{references}
\end{document}